\documentstyle[11pt]{article}
\include{epsf}
\makeatletter
\def\gsim{\compoundrel>\over\sim}

\def\compoundrel#1\over#2{\mathpalette\compoundreL{{#1}\over{#2}}}
\def\compoundreL#1#2{\compoundREL#1#2}
\def\compoundREL#1#2\over#3{\mathrel
         {\vcenter{\hbox{$\m@th\buildrel{#1#2}\over{#1#3}$}}}}
\makeatother
\begin{document}
\setcounter{page}{0}
\thispagestyle{empty}
\begin{flushright}
GUTPA/95/10/3
\end{flushright}
\vskip .1in
\begin{center}
{\Large\bf The Problem of Quark and Lepton Masses
\footnote{Research partially funded by the UK
 Particle Physics and Astronomy Research Council.}}
\vskip 1.0cm
\large
{\bf C. D. Froggatt}
\vskip .4cm
{ \em Department of Physics and Astronomy,
\\University of Glasgow
\\Glasgow G12 8QQ, U.K.}
\end{center}
\section*{ }
\begin{center}
{\Large\bf Abstract}
\end{center}
\large
Different approaches to the fermion mass problem are reviewed. We
illustrate these approaches by summarizing recent developments in models of
quark and lepton mass matrices. Dynamical calculations of the top quark
mass are discussed, based on (a) infrared quasi-fixed points of the
renormalisation group equations, and (b) the multiple point criticality
principle in the pure Standard Model. We also consider Yukawa unification and
mass matrix texture. Models with approximately conserved gauged chiral
flavour charges beyond the Standard Model are shown to naturally give a
fermion mass hierarchy.

\vskip 4.0cm
{\em To be published in the Proceedings of the Seventh Lomonosov
Conference on Elementary Particle Physics, Moscow, Russia,
24 - 30  August 1995.}
\pagebreak
\normalsize
\title{The Problem of Quark and Lepton Masses}
\author{ C. D. Froggatt \\ {\em Department of Physics and
Astronomy,} \\ {\em University of Glasgow,} \\ {\em Glasgow G12 8 QQ,} \\
{\em Scotland, U.K.} \\}
\date{Email address: c.froggatt@physics.gla.ac.uk}
\maketitle

\begin{abstract}
Different approaches to the fermion mass problem are reviewed. We
illustrate these approaches by summarizing recent developments in models of
quark and lepton mass matrices. Dynamical calculations of the top quark
mass are discussed, based on (a) infrared quasi-fixed points of the
renormalisation group equations, and (b) the multiple point criticality
principle in the pure Standard Model. We also consider Yukawa unification and
mass matrix texture. Models with approximately conserved gauged chiral
flavour charges beyond the Standard Model are shown to naturally give a
fermion mass hierarchy.
\end{abstract}
\vspace{0.8cm}
\section{Introduction}
The explanation of the fermion mass and mixing hierarchies and the
three generation structure
of the Standard Model (SM)  constitutes the most important unresolved problem
in particle physics. We shall discuss recent developments in three of the
approaches to this problem:
\begin{enumerate}
\item The dynamical determination of the top quark mass.
\item Mass matrix ans\"{a}tze and texture zeroes.
\item Chiral flavour symmetries and the fermion mass hierarchy..
\end{enumerate}
Neutrino masses, if non-zero, have a different origin to those of the quarks
and charged leptons; we do not have time here to discuss recent applications
of the so-called see-saw mechanism, which seems the most natural way to
generate neutrino masses.

\section{Dynamical Top Quark Mass}
There is presently a lively interest \cite{nambu,dudas,kounnas,smtop} in
determining the top quark mass $m_t$ (or more generally third generation
masses) dynamically. Most of the discussed models lead to the top quark
running Yukawa coupling constant $g_t(\mu)$
being attracted to its infra-red quasi-fixed point value. We have
very recently pointed out \cite{smtop} that the top quark
(and Higgs) mass can be calculated within the pure SM,
assuming the multiple point criticality principle. We now
discuss these two possibilities.

\subsection{Top Mass as a Renormalisation Group Fixed Point}
The idea that some of the properties of the quark-lepton mass spectrum
might be determined dynamically as infrared fixed point values of the
renormalisation group equations (RGE) is quite old \cite{fn1,pendleton,hill}.
In practice one finds an effective infrared stable quasifixed point behaviour
for the SM quark running Yukawa coupling constant RGE at the scale
$\mu \simeq m_t$, where the QCD gauge coupling constant $g_3(\mu)$
is slowly varying. The quasifixed point prediction of the top quark mass
is based on two assumptions: (a) the perturbative SM is valid up to
some high (e.g. GUT or Planck) energy scale
$M_{X} \simeq 10^{15} - 10^{19}$ GeV, and (b) the top Yukawa coupling
constant is large at the high scale $g_{t}(M_{X}) \gsim 1$.
The nonlinearity of the RGE then strongly focuses $g_{t}(\mu)$ at the
electroweak scale to its quasifixed point value. We note that
while there is a rapid convergence to the top Yukawa coupling fixed
point value from above, the approach from below is much more gradual.
The RGE for the Higgs
self-coupling $\lambda(\mu)$ similarly focuses $\lambda(\mu)$ towards a
quasifixed point value, leading to the SM fixed point predictions \cite{hill}
for the running top quark and Higgs masses:
\begin{equation}
m_{t} \simeq 225\ \mbox{GeV} \quad m_{H} \simeq 250\ \mbox{GeV}
\end{equation}
Unfortunately these predictions are inconsistent with the CDF and D0
results \cite{CDF}, which require a
running top mass \mbox{$m_{t} \simeq 170 \pm 12$ GeV}.

However the fixed point top Yukawa coupling is reduced by 15\%
in the Minimal Supersymmetric
Standard model (MSSM), with supersymmetry breaking at the electroweak scale
or TeV scale, due to the contribution of the supersymmetric partners to
the RGE. Also the top quark couples to just one of the two Higgs doublets in
the MSSM, which has a VEV of \mbox{$v_{2} = (174\ \mbox{Gev})\sin\beta$},
leading to the MSSM fixed point prediction
for the running top quark mass
\cite{barger}:
\begin{equation}
m_{t}(m_{t}) \simeq (190\ \mbox{Gev})\sin\beta
\end{equation}
which is remarkably close to the CDF and D0 results for
\mbox{$\tan\beta > 1$}.
%

For large $\tan\beta$ it is possible to have a bottom quark Yukawa coupling
satisfying \mbox{$g_{b}(M_{X}) \gsim 1$} which then approaches an infrared
quasifixed point and is no longer negligible in the RGE for $g_{t}(\mu)$.
Indeed with
$\tan\beta \simeq m_{t}(m_{t})/m_{b}(m_{t}) \simeq 60$
we can trade the mystery of the top to bottom quark mass ratio
for that of a
hierarchy of vacuum expectation values, \mbox{$v_{2}/v_{1} \simeq
m_{t}(m_{t})/m_{b}(m_{t})$},
 and have all the third generation Yukawa coupling constants large:
\begin{equation}
g_{t}(M_{X}) \gsim 1 \quad g_{b}(M_{X}) \gsim 1 \quad g_{\tau}(M_{X}) \gsim 1
\label{tbtaufp}
\end{equation}
Then $m_{t}$, $m_{b}$ and \mbox{$R = m_{b}/m_{\tau}$} all approach infrared
quasifixed point
values compatible with experiment \cite{fkm}. This large $\tan\beta$
scenario is consistent with the idea of Yukawa unification \cite{anan}:
\begin{equation}
g_{t}(M_{X}) = g_{b}(M_{X}) = g_{\tau}(M_{X}) = g_{G}
\label{yukun}
\end{equation}
as occurs in the SO(10) SUSY-GUT model with
the two MSSM Higgs doublets in a single {\bf 10} irreducible representation
and $g_{G} \gsim 1$ ensures fixed point behaviour.
However it should be noted that the equality in Eq.~(\ref{yukun}) is not
necessary, since the weaker
assumption of large third generation Yukawa couplings, Eq.~(\ref{tbtaufp}),
is sufficient for the fixed point dynamics to predict \cite{fkm}
the running masses
$m_{t} \simeq 180 \ \mbox{GeV}$, $m_{b} \simeq 4.1 \ \mbox{GeV}$ and
$m_{\tau} \simeq 1.8 \ \mbox{GeV}$ in the large $\tan\beta$ scenario.
Also the lightest Higgs particle mass is predicted to be
$m_{h^0} \simeq 120 \
\mbox{GeV}$ (for a top squark mass of order  \mbox{1 TeV}).

The origin of the large value of \mbox{$\tan\beta$} is of course a puzzle,
which
must be solved before the large \mbox{$\tan\beta$} scenario can be said to
explain the large \mbox{$m_{t}/m_{b}$} ratio. It is possible to introduce
approximate symmetries \cite{anderson,hall} of the Higgs potential which
ensure a hierarchy of vacuum expectation values - a Peccei-Quinn symmetry and
a continuous $\cal R$ symmetry have been used. However these symmetries
are inconsistent with the popular
scenario of universal soft SUSY breaking mass parameters at the unification
scale and radiative electroweak symmetry breaking \cite{carena}.
Also, in the large $\tan\beta$ scenario, SUSY radiative corrections to $m_{b}$
are generically large: the bottom quark mass gets a contribution proportional
to $v_{2}$ from some one-loop
diagrams with internal superpartners, such as top squark-charged Higgsino
exchange , whereas its tree level mass is proportional to
$v_{1} = v_{2}/\tan\beta$. Consequently these loop diagrams give a
fractional correction \mbox{$\delta m_{b}/m_{b}$} to the bottom quark mass
proportional to $\tan\beta$ and generically of order unity
\cite{hall,carena}. The presence of
the above-mentioned Peccei-Quinn and $\cal R$ symmetries and the associated
hierarchical SUSY spectrum (with the squarks much heavier than
the gauginos and Higgsinos) would protect $m_{b}$ from large radiative
corrections, by providing a suppression factor in the loop diagrams and
giving \mbox{$\delta m_{b}/m_{b} \ll 1$}.
However, in the absence of
experimental information on the superpartner spectrum, the predictions of the
third generation quark-lepton masses in the large $\tan\beta$ scenario
must, unfortunately, be considered unreliable.

\subsection{Criticality and the Standard Model}
Here we consider the idea \cite{glasgowbrioni} that
Nature should choose coupling constant
values such that several ``phases'' can coexist, in a very similar way to
the stable coexistence of ice, water and vapour (in a thermos
flask for example) in a mixture with fixed energy and number of molecules.
The application of this so-called multiple point criticality principle
to the determination of the top quark Yukawa coupling constant requires
the SM (renormalisation group improved) effective Higgs potential to have
coexisting vacua, which means degenerate minima:
$V_{eff}(\phi_{min\; 1}) = V_{eff}(\phi_{min \; 2})$. The important point
for us, in the analogy of the ice, water and vapour system, is that the
choice of the fixed extensive variables, such as energy, the number of
moles and the volume, can very easily be such that a mixture must occur.
In that case then the temperature and pressure (i.\ e.\ the intensive
quantities) take very specific values, namely the values at the triple
point, without any finetuning. We stress that this phenomenon of thus
getting specific intensive quantitities is only {\em likely} to happen
for stongly first order phase transitions, for which the interval of
values for the extensive variables that can only be realised as an
inhomogeneous mixture of phases is rather large.

In the SM, the top quark Yukawa
coupling and the Higgs self coupling correspond to intensive quantities
like temperature and pressure. If these couplings are to be determined
by the criticality condition, the two phases corresponding to
the two effective
Higgs field potential minima should have some ``extensive quantity'',
such as $\int d^4x |\phi(x)|^2 $, deviating ``strongly'' from
phase to phase. If, as we shall assume, Planck units reflect the fundamental
physics it would be natural to interpret this strongly first order
transition requirement to mean that, in Planck units, the extensive
variable densities  $\frac{\int d^4x |\phi(x)|^2}{  \int d^4x }$ = $<|\phi|^2>$
for the two vacua should differ by a quantity of order unity.
Phenomenologically we know that
for the vacuum 1 in which we live, $<\phi>_{vacuum\; 1} = 246$ GeV
and thus we should really expect $<\phi>_{vacuum \; 2}$
in the other phase just to be of Planck order of magnitude.
In vacuum 2 the $\phi^4$ term
will a priori strongly dominate the $ \phi^2$
term. So we
basically get the degeneracy to mean that, at the vacuum 2 minimum,
the effective coefficient $\lambda(\phi_{vacuum\; 2})$ must be
zero with high accuracy. At the same $\phi$-value the derivative
of the renormalisation group improved
effective potential $V_{eff}(\phi)$ should be zero because
it has a minimum there. Thus at the second minimum the beta-function
$\beta_{\lambda}$ vanishes as well as $\lambda(\phi)$.

\begin{figure}[t]
\leavevmode
\centerline{
\epsfxsize=6.75cm
\epsfbox{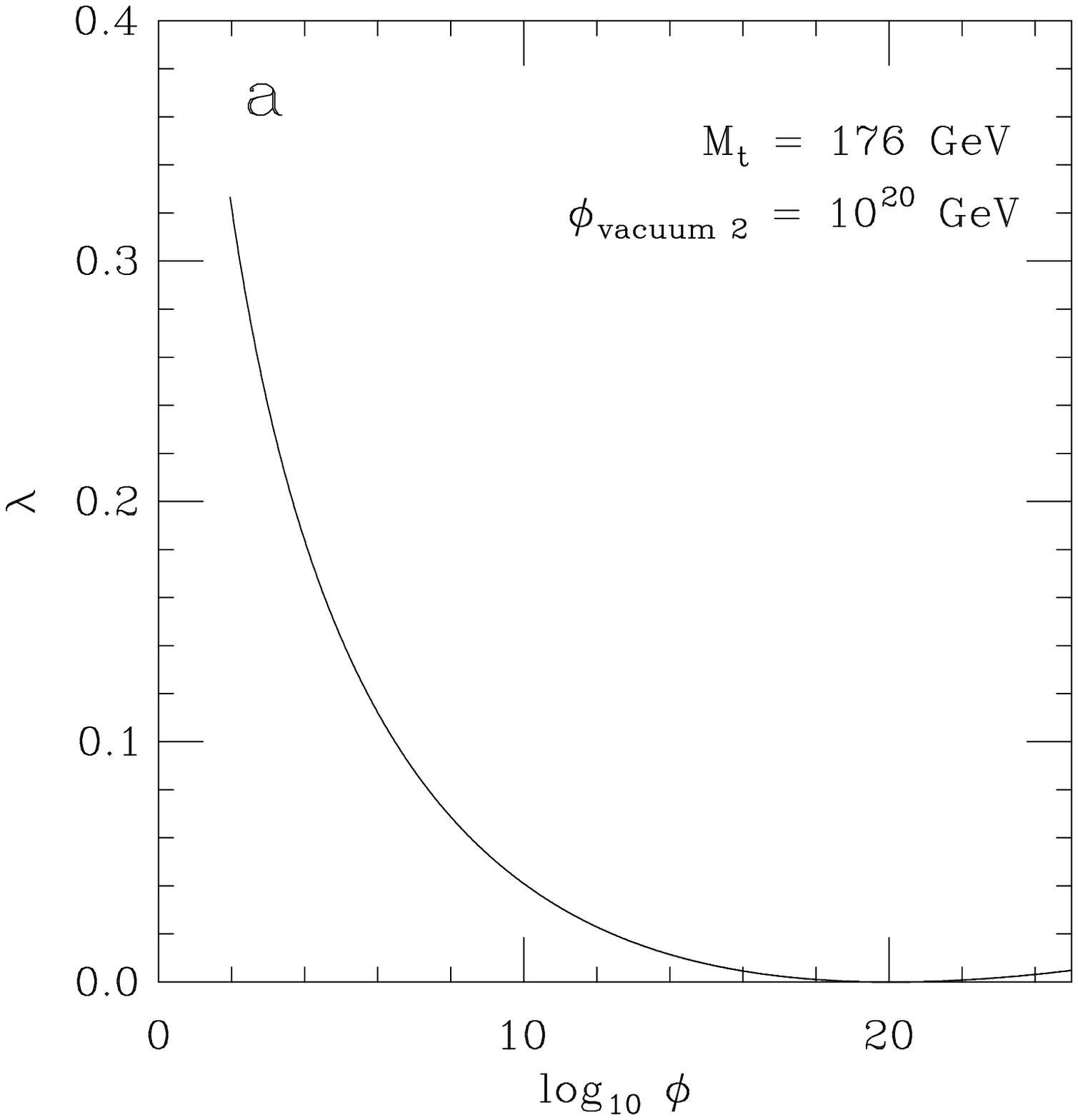}
\epsfxsize=6.75cm
\epsfbox{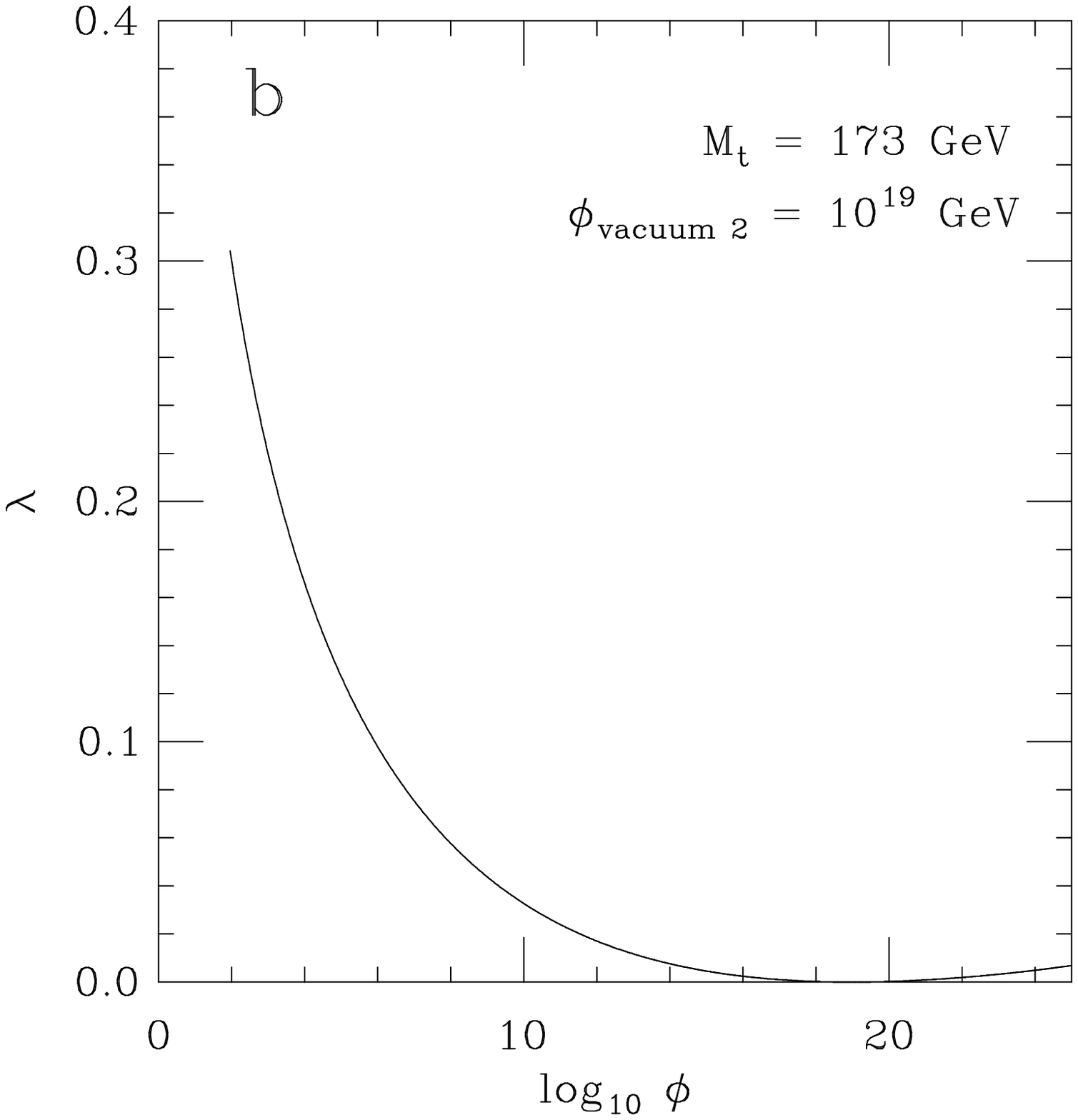}
}
\caption{Plot of $\lambda$ as a function of the scale of the Higgs field
$\phi$ for degenerate vacua with the second Higgs VEV at the scale
(a) $\phi_{vacuum \; 2} = 10^{20}$ GeV and
(b) $\phi_{vacuum \; 2} = 10^{19}$ GeV.
We formally apply the second order SM renormalisation group equations up to
a scale of $10^{25}$ GeV.}
\label{figure}
\end{figure}

We use the renormalisation group to relate the couplings
at the scale of vacuum 2, i.e. at $\mu= \phi_{vacuum\; 2}$, to their values
at the scale of the masses themselves, or roughly at the electroweak scale
$\mu \approx \phi_{vacuum \; 1}$.
Figure \ref{figure}
shows the running $\lambda(\phi)$ as a function of $\log(\phi)$ computed
for two values of $\phi_{vacuum\; 2}$ (where we impose the conditions
$\beta_{\lambda} = \lambda = 0$) near the Planck scale
$M_{Planck} \simeq 2 \times 10^{19}$ GeV.
Combining the
uncertainty from the Planck scale only being known in order of
magnitude and the $\alpha_{QCD}(M_Z) = 0.117 \pm 0.006$ uncertainty
with the calculational uncertainty ,
we get our predicted combination of top and Higgs pole masses:
\begin{equation}
M_{t} = 173 \pm 4\ \mbox{GeV} \quad M_{H} = 135 \pm 9\ \mbox{GeV}.
\end{equation}

\section{Ans\"{a}tze and Mass Matrix Texture}
The best known ansatz for the quark mass matrices
is due to Fritzsch \cite{fritzsch}:
\begin{equation}
M_U =\pmatrix{0  		& C   		& 0\cr
		      C  		& 0   		& B\cr
		      0  		& B   		& A\cr}
\qquad
M_D =\pmatrix{0  		& C^\prime  & 0\cr
		      C^\prime 	& 0   		& B^\prime\cr
		      0  		& B^\prime  & A^\prime\cr}
\end{equation}
where it is necessary to {\em assume}:
$|A| \gg |B| \gg |C|$, $|A^\prime| \gg |B^\prime| \gg |C^\prime|$
in order to obtain a good fermion mass hierarchy.
However, in addition to predicting a
generalised version of the relation
$\theta_c\simeq\sqrt{\frac{m_d}{m_s}}$ for the Cabibbo angle,
which originally motivated the ansatz, it predicts the relationship:
\begin{equation}
|V_{cb}| \simeq
\left| \sqrt{\frac{m_{s}}{m_{b}}} -
e^{-i\phi_{2}}\sqrt{\frac{m_{c}}{m_{t}}} \right|
\label{fritzsch2}
\end{equation}
which cannot be satisfied with a top quark mass \mbox{$m_{t} > 100$ GeV}
\cite{gilman}. Consistency with experiment can be restored by, for example,
introducing a non-zero 22 mass matrix element \cite{xing}. In fact a
systematic analysis \cite{texture} of {\em symmetric} quark mass matrices
with 5 or 6 ``texture'' zeros at the SUSY-GUT scale has been made, yielding
5 ans\"{a}tze consistent with experiment.
Recently ans\"{a}tze incorporating the Georgi-Jarlskog
\cite{georgijarlskog} SUSY-GUT mass relations between leptons and quarks,
$m_{b}(M_X) = m_{\tau}(M_{X})$, $m_{s}(M_X) = m_{\mu}(M_{X})$/3
and $m_{d}(M_X) = 3m_{e}(M_{X})$, have been studied. In particular a
systematic analysis of fermion mass matrices in SO(10) SUSY-GUT models
\cite{anderson,raby} has been made in terms of 4 effective operators.
A scan of millions of operators leads to just 9
solutions consistent with experiment of the form:
\begin{equation}
Y_u =\pmatrix{0               & \frac{-1}{27}C   & 0\cr
		      \frac{-1}{27}C  & 0                & x_{u}'B\cr
		      0               & x_{u}B           & A\cr}
\,
Y_d =\pmatrix{0				  & C				 & 0\cr
			  C				  & Ee^{i\phi}		 & x_{d}'B\cr
			  0				  & x_{d}B			 & A\cr}
\,
Y_l =\pmatrix{0				  & C				 & 0\cr
			  C				  & 3Ee^{i\phi}		 & x_{l}'B\cr
			  0				  & x_{l}B			 & A\cr}
\label{hallraby}
\end{equation}
 For each of the 9 models
the Clebschs $x_{i}$ and $x_{i}'$ have fixed values and
the Yukawa coupling matrices $Y_i$ depend on 6 free parameters: A, B, C, E,
$\phi$ and $\tan\beta$. Each solution has Yukawa unification and
gives 8 predictions consistent with the data.


\section{Chiral Flavour Symmetry and the Mass Hierarchy}
It is natural \cite{fn1} to interpret the fermion mass hierarchy in
terms of partially conserved chiral quantum numbers beyond those of
the SM gauge group. Mass matrix elements are then suppressed by
powers of a symmetry breaking parameter, which may be thought of
as the ratio of the new chiral symmetry breaking scale to the
fundamental scale of the theory. The degree of forbiddenness of a mass
matrix element is then determined by the quantum number difference between
the left- and right-handed SM Weyl states under consideration and the
assumed superheavy fermion spectrum. For example
the four effective operators in the ansatz of Eq. (\ref{hallraby}) can each
be associated with a unique tree diagram, by assigning
an approximately conserved global $U(1)_f$ flavour charge appropriately to
the quarks, leptons and the superheavy states, which are presumed
to belong to vector-like SO(10) {\bf16} + {$\bf \overline{16}$}
representations. The required parameter hierarchy $A \gg B$, $E \gg C$
is naturally obtained in this way and, in particular, the texture zeros
reflect the assumed absence of superheavy fermion states which could
mediate the transition between the corresponding Weyl states.

We now turn to models in which the chiral flavour charges are part of the
extended gauge group. The values of the chiral charges are
then strongly constrained by the anomaly conditions for the gauge theory.
It will also be assumed that any superheavy state needed to mediate a symmetry
breaking transition exists, so that the results
are insensitive to the details of the superheavy spectrum.
The aim in these models is to reproduce all quark-lepton masses
and mixing angles within a factor of 2 or 3.

Ibanez and Ross \cite{ibanezross} have
constructed an anomaly free \mbox{$MSSM \times U(1)_f$} model. The
$U(1)_f$ charges assigned to the quarks and leptons
generate Yukawa matrices of the following form:
\begin{equation}
Y_u \simeq \pmatrix{\epsilon^8  	 & \epsilon^3    	& \epsilon^4\cr
		      		\epsilon^3  	 & \epsilon^2   	& \epsilon\cr
		      		\epsilon^4  	 & \epsilon   		& 1\cr}
\quad
Y_d \simeq \pmatrix{\bar{\epsilon}^8 & \bar{\epsilon}^3	& \bar{\epsilon}^4\cr
		      		\bar{\epsilon}^3 & \bar{\epsilon}^2 & \bar{\epsilon}\cr
		      		\bar{\epsilon}^4 & \bar{\epsilon}	& 1\cr}
\quad
Y_l \simeq \pmatrix{\bar{\epsilon}^5 & \bar{\epsilon}^3 & 0\cr
		      		\bar{\epsilon}^3 & \bar{\epsilon}  	& 0\cr
		      		0  				 & 0   				& 1\cr}
\end{equation}
which are symmetric up to factors of order unity.
The correct order of magnitude for all the masses and mixing angles are
obtained by fitting $\epsilon$, $\bar{\epsilon}$ and $\tan\beta$.
This is a large $\tan\beta \simeq m_t/m_b$ model, but not necessarily
having exact Yukawa unification.
The $U(1)_f$ symmetry is spontaneously broken by two
Higgs singlets, $\theta$ and $\bar{\theta}$, having $U(1)_f$
charges $+1$ and $-1$ respectively and equal vacuum expectation values.
The $U(1)_f^2 U(1)_Y$ gauge anomaly vanishes.
The $U(1)_f^3$ anomaly  and the mixed $U(1)_f$ gravitational anomaly are
cancelled against unspecified spectator particles neutral under the SM group.
However cancellation
of the mixed $SU(3)^2 U(1)_f$, $SU(2)^2 U(1)_f$ and $U(1)_Y^2 U(1)_f$
anomalies is only possible in the
context of superstring theories via the Green-Schwarz mechanism
\cite{greenschwarz} with $\sin^2\theta_W = 3/8$.
Consequently the $U(1)_f$ symmetry is spontaneously broken slightly
below the string scale.

A number of generalisations of this model has been considered during the
last year. By using non-symmetric mass matrices an anomaly free model
has been constructed \cite{pokorski} without the need for the
Green-Schwarz mechanism. Models have also
been considered \cite{pokorski,binetruyramond},
in which the $U(1)_f$ symmetry is broken by just one
chiral singlet field $\theta$ having a $U(1)_f$
charge, say, $-1$. It then follows,
from the holomorphicity of the superpotential, that only positive
$U(1)_f$ charge differences between left and right handed Weyl states
can be balanced by $\theta$ tadpoles. Consequently mass matrix elements
corresponding to negative $U(1)_f$ charge differences have texture
zeros \cite{leurer}. Furthermore if the two
Higgs doublet fields carry $U(1)_f$ charges that do not add up to zero,
the $\mu H_1H_2$ term is forbidden in the superpotential \cite{nir}.
Finally we remark that in effective superstring theories the role
of the $U(1)_f$ symmetry can be played by modular symmetry \cite{dudas},
with the $U(1)_f$ charges replaced by the modular weights of the
fermion fields.

\small

\end{document}